\PassOptionsToPackage{bookmarks={false}}{hyperref} 
\documentclass[sigconf]{acmart} 

\copyrightyear{2020}
\acmYear{2020}
\setcopyright{acmcopyright}
\acmConference[ICSE-SEET'20]{Software Engineering Education and Training}{May
23--29, 2020}{Seoul, Republic of Korea}
\acmBooktitle{Software Engineering Education and Training (ICSE-SEET'20), May
23--29, 2020, Seoul, Republic of Korea}
\acmPrice{15.00}
\acmDOI{10.1145/3377814.3381710}
\acmISBN{978-1-4503-7124-7/20/05}

\usepackage{enumerate}
\usepackage{url}
\usepackage{balance}
\usepackage{enumitem}
\usepackage{soul} 
\usepackage{booktabs} 
\usepackage{pgfplots, filecontents} 

\usepackage{pgfplots} 
\pgfplotsset{compat=1.8} 
\usepgfplotslibrary{statistics} 

\usepackage{tikz}
\usetikzlibrary{matrix,positioning}

\usepackage{ifthen}

\usepackage{subfig} 

\usepackage{amsmath, amsthm, amssymb}
\newtheorem{prop}{Proposition}
\usepackage[export]{adjustbox} 

\usepackage{xspace}

\newcommand{\eg}{\emph{e.g.,}\xspace}
\newcommand{\etc}{etc.\xspace}
\newcommand{\etal}{\emph{et~al.}\xspace} 

\newcommand{\descStep}[2]{\noindent \textbf{#1: } #2}
\newcommand{\descIStep}[2]{\noindent \emph{#1: } #2}

\newlist{learningObjectives}{enumerate}{1}

\setlist[learningObjectives,1]{label={},noitemsep, leftmargin=.13in}%

\newcommand{\LearningObjective}[3]{#1: #2 (#3)}

\newcommand{\ALLURL}{\url{http://all.rit.edu}} 
\newcommand{\TotalStudentsEvaluation}{276~} 

\begin{document}

\title[Impact of Experiential Learning in Computing Accessibility Education]{Presenting and Evaluating the Impact of Experiential Learning in Computing Accessibility Education} 



\author{Weishi Shi}
\affiliation{%
   \institution{Rochester Institute of Technology}
   \city{Rochester}
   \state{NY}
   \country{USA}}
\email{ws7586@rit.edu}

\author{Samuel Malachowsky}
\affiliation{%
   \institution{Rochester Institute of Technology}
   \city{Rochester}
   \state{NY}
   \country{USA}}
\email{samvse@rit.edu}

\author{Yasmine El-Glaly}
\affiliation{%
   \institution{Rochester Institute of Technology}
   \city{Rochester}
   \state{NY}
   \country{USA}}
\email{ynevse@rit.edu}

\author{Qi Yu}
\affiliation{%
   \institution{Rochester Institute of Technology}
   \city{Rochester}
   \state{NY}
   \country{USA}}
\email{qyuvks@rit.edu}

\author{Daniel E. Krutz}
\affiliation{%
   \institution{Rochester Institute of Technology}
   \city{Rochester}
   \state{NY}
   \country{USA}}
\email{dxkvse@rit.edu}





\begin{abstract}


Studies indicate that much of the software created today is not accessible to all users, indicating that developers don't see the need to devote sufficient resources to creating accessible software. Compounding this problem, there is a lack of robust, easily adoptable educational accessibility material available to instructors for inclusion in their curricula. To address these issues, we have created five \emph{Accessibility Learning Labs} (ALL) using an experiential learning structure. The labs are designed to educate and create awareness of accessibility needs in computing. The labs enable easy classroom integration by providing instructors with complete educational materials including lecture slides, activities, and quizzes. The labs are hosted on our servers and require only a browser to be utilized.


To demonstrate the benefit of our material and the potential benefits of our experiential lab format with empathy-creating material, we conducted a study involving \TotalStudentsEvaluation students in ten sections of an introductory computing course. Our findings include: (I) The demonstrated potential of the proposed experiential learning format and labs are effective in motivating and educating students about the importance of accessibility (II) The labs are effective in informing students about foundational accessibility topics (III) Empathy-creating material is demonstrated to be a beneficial component in computing accessibility education, supporting students in placing a higher value on the importance of creating accessible software. Created labs and project materials are publicly available on the project website: \ALLURL 




\end{abstract}

%
%



\maketitle

\section{Introduction}
Approximately 15\% of the world population has a disability~\cite{WHO_Disability_url}, but much of the software created today is inaccessible to people with visual, cognitive, hearing, dexterity, and other disabilities~\cite{wash_designing_url, lazar2007frustrates, Calvo:2016:BWC:3019943.3019955, Gonçalves2013, Hanson:2013:PWA:2435215.2435217, disability2004web}.~
Addressing this problem necessitates an accessibility-literate workforce that not only understands how to create accessible software, but also recognizes the impact inaccessible software can have on many users. Although accessibility is a vital computing topic, it is often excluded from formal undergraduate education~\cite{keith2009design, teacha11y_url}. Additionally, research indicates that computing instructors have the desire to integrate accessibility-related topics in their courses, however they frequently lack access to teaching materials to use in their courses~\cite{Shinohara:2018:TAS:3159450.3159484, Kawas:2019:TAD:3287324.3287399}. 

To fill the current void in accessibility education, we created a \emph{comprehensive collection of laboratory activities to benefit accessibility education.} These labs are collectively referred to as the \emph{Accessibility Learning Labs} (ALL), and have the primary goals of creating student awareness of the need to create accessible software and to inform students about foundational accessibility concepts. No special software is required to use any portion of the labs since they are web-based and hosted on our servers, requiring only a browser and an internet connection for usage. The labs are easily integrated into existing introductory computing courses such as Computer Science I \& II (CS1 \& CS2) due to their easy-to-adopt, self-contained nature. Each lab has a designated difficulty rating (Introductory, Medium, Advanced) to maximize impact regardless of course levels, specialization and student experience. 

Each lab addresses at least one foundational computing accessibility topic and contains: I) Relevant background information on the examined topic, II) An example application containing the accessibility problem, III) A process to emulate the accessibility problem (as closely as possible), IV) Details about how to repair the problem from a technical perspective, and V) Information from people about how this encountered accessibility issue has impacted their life. As an example, the color blindness (deuteranope) lab includes information about the condition, an on-screen simulation, a way to solve the issue, and a video where a user with this condition discusses how inaccessible apps have impacted their computing experiences.

A key component of this effort is creating empathy for users with disabilities, motivating students to create accessible software for these users by preparing students with technologies and software design methods that address those concerns. 
Our labs contain empathy-creating `supplementary material' designed to demonstrate the necessity of creating accessible software.



To demonstrate the effectiveness of our labs and their experiential learning format, we evaluated them in ten sections of CS2 that included a total of~\TotalStudentsEvaluation students and found that: (I) The demonstrated potential of the proposed experiential learning format and labs are effective in motivating students about the importance of accessibility, (II) The proposed material is effective in informing students about foundational accessibility topics, and (III) Empathy-creating material is demonstrated to be a beneficial component in accessibility education, supporting students in placing a higher value on the importance of creating accessible software. 




To summarize, this work makes the following contributions:

\begin{itemize}[noitemsep]

    \item \descStep{Systematic evaluation}{In our analysis, we demonstrate the effectiveness of our proposed experiential learning structure against existing material through the use of our labs. Our findings demonstrate that the proposed labs and their structure are more effective than existing material at increasing student understanding and motivation in creating accessible software.}

    \item \descStep{Experiential accessibility education material}{Our self-contained labs represent the first experiential educational accessibility material that is publicly available, contains all necessary material for complete classroom adoption, and is web-hosted to enable easy adoption. Our five created labs are publicly available on our project website: \ALLURL}

    \item \descStep{Demonstrate importance of empathy-creating materials in accessibility education}{We demonstrate that providing empathy-creating material, such as videos of student-peers with accessibility issues, is beneficial for increasing student understanding for the importance of creating accessible software.}
    



\end{itemize}

The rest of the paper is organized as follows: Section~\ref{sec: Labs} presents the general structure of our created labs, Section~\ref{sec: Evaluation} describes our analysis and discovered results, Section~\ref{sec: Discussion} discusses the discovered results, including limitations to our study and future work, Section~\ref{sec: RelatedWork} presents related works, and Section~\ref{sec: Conclusion} provides a conclusion.





\section{Accessibility Learning Labs}
\label{sec: Labs}



The following subsections will describe the goals of the \emph{Accessibility Learning Labs} (ALL), their components, and addressed accessibility topics.


\subsection{Lab Goals} 
\label{sec: LabGoals}

The educational accessibility learning labs have been systematically developed to achieve the following key goals:

\begin{enumerate}

    \item \descIStep{The labs will not require any special hardware or software}{Only a web-browser will be needed to run the labs, allowing institutions and individual learners that don't have the ability to install special software or have older computers to easily utilize the labs. This will also support classroom inclusion by eliminating classroom computer pre-configuration time for already busy instructors.}



    \item \descIStep{Instructors and students will only require very basic programming/computing skills to utilize the labs}{It is imperative for people with all levels of software development abilities to recognize the importance of creating accessible software. Therefore, the labs do not require any substantial special technical skills or knowledge of any specific programming language. This supports the inclusion of the labs into introductory computing courses that utilize a wide-range of tools and technologies.}

    \item \descIStep{The labs should fit into already crowded foundational computing courses}{Each lab is designed to take approximately 20-60 minutes, and the instructor may select the lab components that they would like to utilize in an \`a-la-carte fashion inside or outside of the classroom. The succinctness of the labs will enable them to fit into courses that are already heavily  time-constrained.}

    \item \descIStep{The labs should include all instructional content}{Each lab should represent a complete educational experience for the student. To support this, labs contain all necessary material required for classroom inclusion. This includes lecture slides, background reading material on the accessibility issue, and how it can be repaired from a technical perspective.}

    \item \descIStep{The labs should demonstrate the need to create accessible software}{A primary goal of the labs is to establish the importance of creating accessible software for students. Each lab will enable students to experience the accessibility issue being addressed (as closely as possible) as well as additional motivating material in the form of written and video testimonials from individuals describing how inaccessible software has had an adverse impact on their computing experiences.}

\end{enumerate}

These goals are important since a primary objective of the labs is to allow the inclusion of accessibility at resource-constrained institutions that might not necessarily have the ability to include accessibility in their courses.


\subsection{Lab Components} 
\label{sec: LabStructure}

Each lab is comprised of several components which are systematically designed to inform and motivate students about the topic of accessibility. These lab components are described below.



\vspace{2mm} \noindent \textbf{Background Instructional Material:} Each lab contains instructional material in several formats. This includes a brief written description (2-4 minutes of reading), lecture slides (.pptx and .pdf format) and background material on the addressed accessibility topic. An Americans with Disabilities Act (ADA)~\cite{ADAVideo_url}-compliant screencast of the lecture slides is available if the instructor would prefer to show the video in class or have the students view the video outside of the classroom. The lecture slides and videos are designed to take approximately 3-5 minutes. The objective is to provide the instructor all necessary materials to include the topic of accessibility in their course and also enable the instructor flexibility to alter any of the material as they see fit. The instructor may also choose to use the material in an \`a-la-carte fashion if they desire.

\vspace{1mm} 
\noindent \textbf{Activity:} Students interact with the experiential activity through their browser. Each lab activity is comprised of the following steps:

\begin{figure*}[h!]
\centering
\subfloat [Inaccessible software since user cannot hear notification and the visual message is not relevant] {\tikz[remember
picture]{\node(1A){\includegraphics[scale=.25]{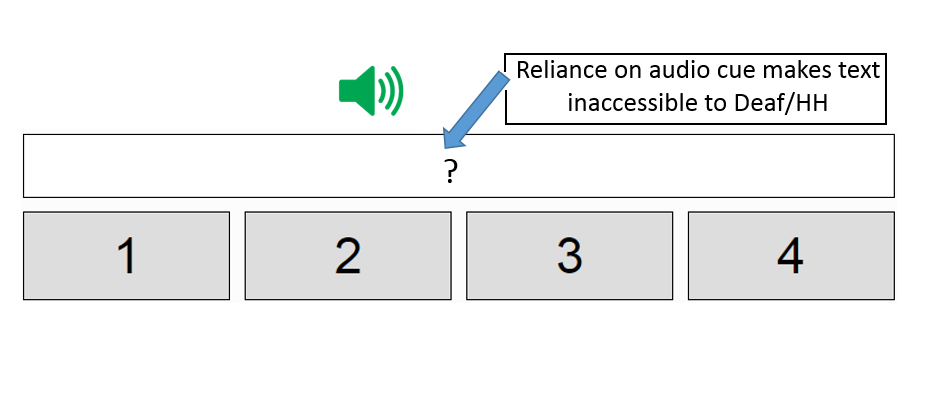}\label{fig:LabFixA}};}}%
\hspace*{.33cm}%
\subfloat [Mock IDE used through browser]{\tikz[remember picture]{\node(2A){\includegraphics[scale=.25]{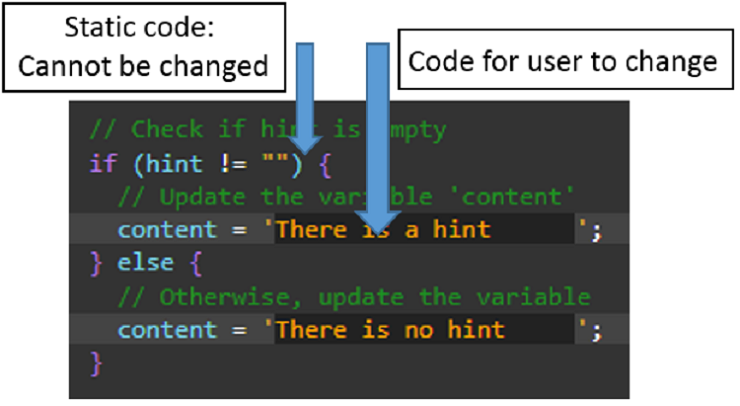}\label{fig:LabFixB}};}}
\hspace*{.33cm}%
\subfloat [Software made more accessible by student adding informative visual message.]{\tikz[remember picture]{\node(3A){\includegraphics[scale=.25]{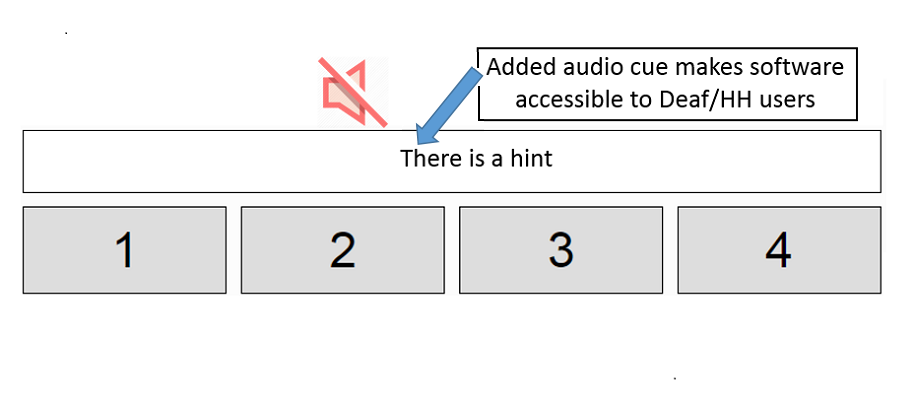}\label{fig:LabFixC}};}}

\caption{Example of student repairing accessibility problem using simulated IDE}

\label{fig:LabFix}

\tikz[overlay,remember picture]{

    \draw[-latex,thick] (1A) -- (1A-|2A.west) node[midway,below,text width=.3cm]{};
    \draw[-latex,thick] (2A) -- (2A-|3A.west) node[midway,below,text width=.3cm]{};

    

}
\end{figure*}


\begin{enumerate}


    \item \descStep{Students interact with software}{Students interact with the software without any accessibility emulation feature, meaning that they will experience the software as a person with typical ability would. Students will then be instructed to perform a simple task in the application. For example, in our colorblindness lab (Lab \#1) students play a game where they need to quickly select a specifically colored circle when it appears.} 

    \item \descStep{Students experience accessibility challenges through an emulation feature}{Each lab contains a feature to emulate the addressed accessibility topic as closely as possible. For example, in Lab \#3 (blindness), the text is blurred to emulate what a user with a visual impairment would experience. The objective is to demonstrate adverse impacts first-hand.} 

    \item \descStep{Details are provided on how to repair the application}{Students are next provided best practices to repair the encountered accessibility issue. This fix varies by lab, and may include using specific colors to make the application more accessible to colorblind users, or properly incorporating `alt' tags for users with screen readers.} 

    \item \descStep{Students repair the accessibility problem}{As shown in Figure~\ref{fig:LabFix}, students repair the accessibility problem using a simulated code editor on the project web application. For example, to ensure that software is accessible to users with hearing impairments, a common best practice will be to not rely solely upon audio cues since this can adversely impact the experience for Deaf/Hard of Hearing users. Participants next repair the application through the use of appropriate text or visual aides in addition to any audio notifications~\cite{UF-Cues, WC3-Cues}. Figure~\ref{fig:LabFixA} demonstrates the inaccessible version of the software, while Figure~\ref{fig:LabFixB} shows the simulated IDE to change the source code. Figure~\ref{fig:LabFixC} demonstrates the repaired, accessible version of the application.}



    \item \descStep{Students use the software with the emulation feature active, but with their modifications in place}{This phase enables the student to experience the impact of their alterations in making the software more accessible and evaluate the impact of their changes. This also instills student confidence that they are capable of making accessible software.}
    

\end{enumerate}

\vspace{1mm} \noindent \textbf{Empathy-Creating Supplementary Material: }Providing students the proper technical knowledge necessary to create accessible software is important, but demonstrating the \emph{importance} of creating accessible software is paramount for motivating students to learn about creating accessible software~\cite{putnam2016best, Putnam:2015:TAL:2700648.2811365, McQuiggan:2008:EEV:1357054.1357291}. Supporting this, each lab contains supplementary awareness creating materials such as discussions by people with the addressed accessibility issue.

An example in the empathy-creating component is a person who is Deaf/Hard of Hearing using transcribed American Sign Language (ASL) to discuss the negative impact of attempting to use software that relies upon audio cues, rendering it inaccessible for users similar to them (Lab \#1). 
The people in these videos are undergraduate students (age 18-22). We believe that students completing the labs will better identify with accessibility challenges encountered by users of their peer age group. This material will be provided in both written and video format through the project website. Proper IRBs were attained prior to creating this material.


\vspace{1mm} \noindent \textbf{Quiz: }For each lab, adopting instructors may request access to a brief quiz (approximately 10 questions) intended to assist them with student summative evaluations. The questions for each quiz are derived from the provided instructional material, and an answer key is provided. Instructors may request access via the project website and after a simple instructor-verification process will be emailed the quiz material. All quiz material was reviewed by our accessibility and instructional design experts to ensure robustness. 




\subsection{Lab Topics}
\label{sec: LabTopics}



Each lab is focused on defined learning objectives (Bloom's Taxonomy)~\cite{bloom1956taxonomy} and is targeted for students in one of three proficiency levels (I) Introductory: Little proficiency in computing, (II) Intermediate: Basic computing proficiency, consistent with foundational computing courses, or (III) Advanced: Medium to high computing proficiency, consistent with upper-level computing courses. The accuracy and appropriateness of each lab was verified by both our internal development team consisting of accessibility and instructional design experts and our project's external advisory board. This advisory board is comprised of a practicing Speech Language Pathologist (SLP) and two accessibility experts from external institutions. An overview of each lab follows. 




\newcommand{\LabHeading}[3]{\vspace{1mm} \noindent \textbf{Lab #1 - #2 (#3): }}



\LabHeading{1}{Using visual cues to make software accessible to \\ Deaf/Hard of Hearing users}{Introductory}This lab serves to introduce the concept of making software accessible to users who are Deaf/Hard of Hearing. This lab involves students playing a game where they are tasked with locating a random, hidden item. Points are awarded for finding the item quickly. An audio cue randomly provides the location of the hidden item, thus enabling the user to identify it sooner with more accuracy and achieve a higher score. The accessibility emulation component involves merely not playing the audio cue, emulating the experience of a person who is Deaf/Hard of Hearing. To make the software more accessible, the student adds a visual cue for the hint, thus making the software more accessible to Deaf/Hard of Hearing users. An example of this feature is shown in Figure~\ref{fig:LabFixA} and Figure~\ref{fig:LabFixC}.

\vspace{2mm} \noindent \underline{Learning Objectives (LO)} - After completion of the lab, students should be able to: \vspace{-1mm}

\begin{learningObjectives}
    \item\LearningObjective{LO1}{Recognize difficulties Deaf/Hard of Hearing individuals may encounter when using inaccessible software}{Comprehension}
    \item\LearningObjective{LO2}{Examine accessibility solutions specific to Deaf/Hard of Hearing challenges}{Analysis}
    \item\LearningObjective{LO3}{Construct a more Deaf/Hard of Hearing accessible version of an existing application}{Synthesis}
    
\end{learningObjectives}






\LabHeading{2}{Making software accessible to users who are colorblind}{Introductory}The primary learning objective of this lab is to inform students about the \emph{Distinguishable Content} accessibility guideline \cite{standards1_url}. This lab introduces the concept of making software accessible to users who are colorblind. Students are presented with a game and are asked to click on specifically-colored circles when they appear. After the first round of the game, a color-blindness emulation feature is activated, which makes the colors appear similar as they would to someone who is colorblind. This makes the task of clicking on specifically defined colors very difficult since most of the colors will now appear indistinguishable from one another. Students are then tasked with using (different) proper colors to make the software accessible to users who are colorblind (\eg Deuteranope). This is accomplished by having the student modify the colors used in the application and then using the software with the colorblindness emulation feature still active to experience the impact of their alterations. 

\vspace{2mm} \noindent \underline{Learning Objectives (LO)} - After completion of the lab, students should be able to: \vspace{-1mm}

\begin{learningObjectives}
    \item\LearningObjective{LO1}{Recognize difficulties that colorblind (Deuteranope) individuals may encounter when using inaccessible software}{Comprehension}
    \item\LearningObjective{LO2}{Examine accessibility solutions specific to colorblindness-related challenges}{Analysis}
    \item\LearningObjective{LO3}{Construct a more colorblind accessible version of an existing application}{Synthesis}
\end{learningObjectives}



\LabHeading{3}{Making software accessible to blind users}{Medium}This lab focuses on demonstrating the importance of creating software that is accessible to users who are blind and the foundational practices that may be incorporated to make the software accessible to these users. This activity contains two primary stages. This first stage involves students interacting with a page and clicking on images of a specific type of animal (\eg cats), which will be a trivial task for seeing students. The next step has the student install a screen-reader add-on (ChromeVox~\cite{ChromeVox_url}) and then perform the same task with an inaccessible version of the page. However, this time the page has a dark box hiding the images on the screen and, since the page is not accessible, the audio information provided by the screen reader provides no value. Because the images do not contain properly informative \emph{alt} tags~\cite{PSU-Alt_url, Wash-Alt_url}, they don't contain useful information for the screen reader. 
The `repair' component of the activity involves students adding informative alt tags to each of the images, thus making the page accessible to blind users who rely on screen readers. Similar to other labs, students will be able to experience the impact of their change through the hosted web application.

The second, more advanced phase involves students identifying the inaccessible portions of a provided web page. Using knowledge gained from the provided accessibility material and lectures, students are shown a page with several features that are inaccessible to users with visual impairments, with some examples being poor contrast, poorly labeled hyperlinks, images for text and poorly structured headings~\cite{W3CWebDesignApplicaitons_url, W3LowVision_url, ADAWebsites_url}. Students are tasked with identifying these inaccessible components and repairing them. We believe that the ability to identify inaccessible components of a web page is an important skill for students to gain experience in, especially for when they are developing new software and modifying legacy applications in the real-world.

\vspace{2mm} \noindent \underline{Learning Objectives (LO)} - After completion of the lab, students should be able to: \vspace{-1mm}

\begin{learningObjectives}
    \item\LearningObjective{LO1}{Recognize difficulties that blind individuals may encounter when using inaccessible software}{Comprehension}
    \item\LearningObjective{LO2}{Examine accessibility solutions specific to blindness-related challenges}{Analysis}
    \item\LearningObjective{LO3}{Construct a more accessible version of an existing application for blind users}{Synthesis}
\end{learningObjectives}



\LabHeading{4}{Introduction to dexterity issues}{Medium}This lab introduces students to the importance of making software accessible to users with dexterity issues. In this lab, students are asked to begin by clicking a small ``go'' button to begin a fictitious task. However, the students will find it difficult to click the button as the sensitivity of the mouse is set to high and the button will move slightly as the mouse approaches it, emulating the experiences of a user with a motor disability. 
The student will be asked to adjust the CSS to make the button larger and to meet the accessibility guidelines, hence making it easier to click and more accessible to users with dexterity issues. 

An additional section has the student complete an `account creation' form using only their keyboard. Forcing the student to only use a keyboard will closely emulate the experiences of a user who is unable to use a mouse. To reach the form, students must traverse through a long navigation bar. Due to the page not being created properly, students will notice that this task is quite inconvenient. This demonstrates the need to include a <main> tag in the HTML so that a user can quickly skip to the main section of a page using a keyboard. Students are tasked with making this correction in the HTML portion of the hosted application. 

The last component of this activity is another form. However, this time the form includes a tooltip that is inaccessible using only a keyboard. The tooltip contains important information for input constraints which is necessary to complete the form. Since the students cannot see the hint from the tooltip, they will not be able to successfully complete the form. Students will then be prompted to fix the CSS in the hosted web application to include tab-index so the tooltip can be accessed. This demonstrates the importance of creating software that is keyboard accessible so that users with dexterity issues, who cannot use a mouse, can still use the application using only a keyboard.

\vspace{2mm} \noindent \underline{Learning Objectives (LO)} - After completion of the lab, students should be able to: \vspace{-1mm}

\begin{learningObjectives}
    \item\LearningObjective{LO1}{Recognize difficulties that individuals with dexterity challenges may encounter when using inaccessible software}{Comprehension}
    \item\LearningObjective{LO2}{Examine accessibility solutions specific to dexterity-related challenges}{Analysis}
    \item\LearningObjective{LO3}{Construct a more accessible version of an existing application for users with dexterity challenges}{Synthesis}
\end{learningObjectives}




\LabHeading{5}{Making software accessible to users with cognitive impairments}{Advanced} This lab will make students aware of the accessibility guidelines for users with cognitive Impairment~\cite{murphy2005accessibility, sevilla2007web, jaramillo2017accessibility}. Some of the cognitive accessibility problems addressed in this lab include too many objects displayed at the same time, lack of logic (consistent actions lead to inconsistent results), small text and rows containing too much text. Some of the covered best practices include minimizing cognitive load, limiting the number of typefaces in the document, and providing regular feedback to users. In this activity, students are provided with a set of pages that are inaccessible to users with cognitive impairment. The students are tasked with identifying and repairing the accessibility problems with these pages. 


\vspace{2mm} \noindent \underline{Learning Objectives (LO)} - After completion of the lab, students should be able to: \vspace{-1mm}

\begin{learningObjectives}
    \item\LearningObjective{LO1}{Recognize difficulties that users with cognitive disabilities may encounter when using inaccessible software}{Comprehension}
    \item\LearningObjective{LO2}{Examine accessibility solutions specific to cognitive-related challenges}{Analysis}
    \item\LearningObjective{LO3}{Construct a more accessible version of an existing application for users with cognitive disabilities}{Synthesis}
\end{learningObjectives}







\subsection{Lab Availability}
\label{sec: LabAvailability}
Users require only an internet connection and web browser (Safari, Chrome, Edge, Firefox) for adoption. Complete lab material including lecture slides, videos, quiz, and activities are publicly available on our project website: \ALLURL



Supplementary ADA-compliant videos are available through our YouTube channel. These include videos of the lectures, the activity being conducted and empathy-creating supplementary material. Also included is a video that Intuit created for our project, where a manager and engineer discuss the necessity of creating accessible software. They also discuss how the ability to create accessible software is an important trait during their hiring process. We have made this video available so that others may show it to their students to further demonstrate the real-world need for creating accessible software, and that it is an important and attractive skill for developers to have from an organizational perspective. The link to our YouTube channel is available on our project website. 


\section{Evaluation}
\label{sec: Evaluation}



\newcommand{\RQA}{How effective are the labs in motivating students about the importance of accessibility?} 
\newcommand{\RQB}{How effective are the labs in informing students about foundational accessibility principles?} 

\newcommand{\RQC}{How impactful are `empathy-creating' materials in accessibility education?} %

Our work addresses the following research questions:

\hangindent=3.6em 
\textbf{RQ1.} \emph{\RQA~}Through an experiment using our material, a statistical analysis demonstrates the positive impact our material has in motivating students on the importance of computing accessibility education.

\hangindent=3.6em 
\textbf{RQ2.} \emph{\RQB~}A statistical analysis demonstrates that our material using our experiential format is more effective in informing students about foundational accessibility principles, while activities containing empathy-creating material have a higher universal positive effect on students.

\hangindent=3.62em 
\textbf{RQ3.} \emph{\RQC~}A t-test demonstrates that additional empathy-creating material can be an important component of accessibility education. These observations support findings from RQ1 and RQ2 about the importance of empathy-creating material in accessibility education.



\subsection{Experimental Design}
\label{sec: ExperimentalDesign}













To evaluate our created material and the potential benefits of their experiential format, we included one of our labs in ten sections of a CS2 course in a conventional classroom format at our university with \TotalStudentsEvaluation students participating. The CS2 course is primarily comprised of Computer Science, Software Engineering, Computing Security, and Computer Engineering majors. The vast majority of students were first year, second semester students. Our first year program does not include formal educational accessibility activities, so this is very likely the first formal accessibility activity that the students participated in at our institution. We also surmise that this is likely the first formal accessibility training that many students will have participated in at any point (an assumption that is supported by survey results later described in this section). 
We selected Lab \#2 (colorblindness) for evaluation due to its introductory proficiency level and appropriateness for our specific offering of CS2.

We created a pre-lab-survey, post-lab-survey, and quiz to evaluate the impact of our material. Survey and quiz questions were developed and reviewed by our instructional design and accessibility experts prior to usage. We used a random number generator to place each of the ten course sections into groups A, B, or C, where the first four selected groups were assigned to Group A and the next three into Group B and C. Four sections were placed into Group A since this would be frequently used as a control group to compare findings against a combination of Group B and Group C.

\begin{itemize}

    \item \descStep{Group A: Control Group}{The control group utilized existing material to instruct students about the addressed accessibility concept. In this evaluation, to cover the topic of colorblindness, we selected material from Mozilla~\cite{MozillaAccessability_url} since it is a well-known resource from an established organization. Students in Group A using this existing material were asked to follow this provided instructional content, which we did not make any alterations to.}

    \item \descStep{Group B: Our labs - \emph{without} `supplementary' material}{This group used our material, except for the `supplementary material' (described in Section~\ref{sec: LabStructure}). The purpose of excluding this supplementary material was to enable the evaluation of its effectiveness for both informing and motivating students about the topic of accessibility.} 

    \item \descStep{Group C: Our labs - \emph{with} `supplementary' material}{This group used all created lab materials, including the empathy-creating content. This group provided the ability to evaluate the empathy-creating material and its effectiveness in accessibility education.}


\end{itemize}



To provide the necessary evaluation data, each of the three groups used the following steps to conduct the activity:


\begin{enumerate}

    \item \descStep{Pre-lab-survey}The pre-lab-survey provided relevant background on the students including their major and year level. This instrument also provided us with a baseline for the student's interest level in accessibility and their belief of the importance of creating accessible software.
    
    \item \descStep{Provide Background materials}{Students are provided background material on the addressed concept. This reading is designed to take approximately 2-5 minutes. For students using our materials (Group B and Group C), they were also shown a brief lecture video which provided instructional material on the examined topic.}

    \item \descStep{Conduct Activity}{Students then conduct the hands-on activity. Group A utilized existing material, while Groups B and C used our created material.}

    \item \descStep{Interact With Supplementary Material (Group C only)}{The students in this group also interacted with our provided supplementary material (described in Section~\ref{sec: LabStructure}).}

    \item \descStep{Quiz}{Students were asked to complete a ten question quiz at the conclusion of the activity. We used an identical quiz for each group and ensured that all quiz questions were covered in both the Mozilla material our material. The inclusion of the quiz not only enabled us to evaluate the knowledge gained by the students, but also the effectiveness of the provided quiz material for the instructors. The quiz questions focused on evaluating the student's comprehension of technical concepts and their understanding of the addressed accessible topic in general. This quiz was created under the guidance of our instructional design and accessibility experts.}


     \item \descStep{Post-lab-survey}{A post-lab-survey contained a majority of questions analogous to the pre-lab-survey questions. This provided us with a comparative mechanism to evaluate existing and created material.}

\end{enumerate}

For each quiz and survey instrument, students were required to login with their university Google account. This provided us with a mechanism to track and correlate results. In accordance with our IRB, we assigned all students an anonymized ID in our database and removed any personally identifiable information.

\subsection{Overview of collected Data}

Table~\ref{table: groupParticpants} provides an overview of the number of students for each group, broken down by major (Computer Science, Software Engineering, Computing Security, Computing Engineering, or Other). Our results only include students who complete all instruments (pre-lab-survey, post-lab-survey, and the quiz). Each evaluation group is a mixture of students from different education years. Table \ref{table: yearTypeDistribution} demonstrates that the vast majority of students (91\%) identified as first year students (our undergraduate degree is a five year program). Additionally, based on the pre-lab-survey results 67\% of students stated that they had `No experience' with the topic of software accessibility.

\begin{table}[h]
\begin{center}
\caption{Students by major for each group}
\label{table: groupParticpants}
  \begin{tabular}{ c | c | c | c | c | c || c } \hline
 \toprule
  \bfseries Group & \bfseries CS & \bfseries SE & \bfseries Security & \bfseries CE  & \bfseries Other & \bfseries Total \\  \midrule 
  
  \bfseries A & 35 & 24 & 29 & 21 & 12 & 121 \\ \hline
  \bfseries B & 27 & 21 & 17 & 9 & 8 & 82 \\ \hline
  \bfseries C & 27 & 19 & 7 & 13 & 7 & 73 \\ \hline \hline
  \bfseries Total & 89 & 64 & 53 & 43 & 27 & \TotalStudentsEvaluation \\ \hline
\bottomrule
   \end{tabular}
  \end{center}
\end{table}


\begin{table}[h]
\begin{center}
\caption{The year type distribution of the students}
\label{table: yearTypeDistribution}
  \begin{tabular}{ c | c | c | c | c | c } \hline
 \toprule
  \bfseries Group & \bfseries Year 1 & \bfseries Year 2 & \bfseries Year 3 & \bfseries Year 4  & \bfseries Year 5  \\  \midrule 
  \bfseries A & 109 & 7 & 2 & 2 & 1  \\ \hline
  \bfseries B & 73 & 8 & 0 & 1 & 0   \\ \hline
  \bfseries C & 70 & 3 & 0 & 0 & 0   \\ \hline \hline
  \bfseries Total & 252 & 18 & 2 & 3 & 1 \\ \hline
\bottomrule
   \end{tabular}
  \end{center}
\end{table}




%


\subsection{Analysis Results}
\label{sec: analysisResults}




\hangindent=2.9em \hangafter=1
\textbf{RQ1.} \emph{\RQA} 

To answer this first research question, we compared Group A (existing material) against Group B \& Group C (created material). The comparison was conducted using these groups as we wanted to determine the impact that our experiential educational format and labs (Group B \& Group C) had in comparison with existing materials (Group A). We used the pre-and post-lab-survey question of ``How important is it for you to create accessible software?'' to determine the impact our material had on this research question. The survey used a Likert scale of low to high importance. We conducted a dependent t-test over the two pairs of scores since each of them is given by a specific student. Let \textbf{pr} and \textbf{po} denote the n-dimension pre-lab-survey and post-lab-survey vectors of scores respectively, the t-scores were then calculated as follows,
\begin{equation}
    t=\frac{\overline{\Delta \textbf{p}}-\mu_0}{\textbf{s}_{\Delta p} \cdot n^{-\frac{1}{2}}}=\frac{\overline{\textbf{pr}}-\overline{\textbf{po}}}{||(\textbf{pr}-\textbf{po})-(\overline{\textbf{pr}}-\overline{\textbf{po}})||_2 \cdot n^{-\frac{1}{2}}}
\end{equation}

where $\overline{\textbf{pr}}$ and $\overline{\textbf{po}}$ are vector means of \textbf{pr} and \textbf{po} respectively. The constant $\mu_0$ is set to zero because 
we state the null-hypothesis $\mathcal{H}_0$ as the expected rating on the importance of the topic does not change significantly from post-lab-survey to pre-lab-survey. Generally speaking, an activity that significantly impacts student opinion will result in small P-values. Table \ref{table: t-test1} summarizes the p-values from the t-tests.

\begin{table}[h!]
\begin{center}
\caption{P-values of the t-tests for RQ1}
\label{table: t-test1}
  \begin{tabular}{ c | c | c | c || c} \hline
 \toprule
  \bfseries Group & \bfseries  $\overline{\textbf{pr}}$ &  \bfseries $\overline{\textbf{po}}$ & \bfseries $\overline{\Delta \textbf{p}}$  & \bfseries P-value  \\  \midrule 
  \bfseries A & 3.69 & 3.85 & +0.17 & 0.04\\ \hline
  \bfseries B & 3.93 & 4.05 & +0.12 & 0.13  \\ \hline
  \bfseries C & 3.62 & 3.99 & +0.37 & $1e^{-4}$\\ \hline
\bottomrule
   \end{tabular}
  \end{center}
\end{table}

The t-test demonstrates that all three groups improve the students' mean score. However, the P-value suggests that the improvement of Group B is not significant at the 95\% confidence level. Although Group A and Group C both have significant improvement, the absolute improvement of Group C, $\overline{\Delta \textbf{p}}$, is twice as much as that of Group A. This indicates that Group C has a higher positive impact on students than Group A. What is more, the extremely small P-value of Group C makes such conclusion consistent with the data we observed even at the confidence level of 99.9\%. So we conclude from the t-test that Group C can effectively reinforce the students' opinions on how important the topic is. We believe such impact is
beneficial under the assumption that students pay more attention and expend more efforts on topics that they believe to be important~\cite{harackiewicz2010importance, harackiewicz2016interest}.

\begin{figure}[h]
 \begin{center}
  \begin{tikzpicture}[scale=.8]
        \begin{axis}[
            view={0}{90},   
            xlabel=Post-Lab-Survey Scores,
            ylabel=Pre-Lab-Survey Scores,
            colorbar,
            colorbar style={
                title=,
                yticklabel style={
                    /pgf/number format/.cd,
                    fixed,
                    precision=1,
                    fixed zerofill,
                },
            },
            title=Group A: How important to create accessible software,
            %
            enlargelimits=false,
            axis on top,
            point meta min=0,
            point meta max=27,
            %
        ]


            \addplot [matrix plot*,
            point meta=explicit,
            nodes near coords,
            every node near coord/.style={white, yshift=-6pt}] file [meta=index 2] {HM-AJ.dat};
        \end{axis}
    \end{tikzpicture}
    	\caption{Score transition heatmap of Group A. The number
in i-th row and j-th column indicates how many students
change their survey score from i to j in Group A (existing material).}
  	\label{fig:HM-AJ}
\end{center}
\end{figure}
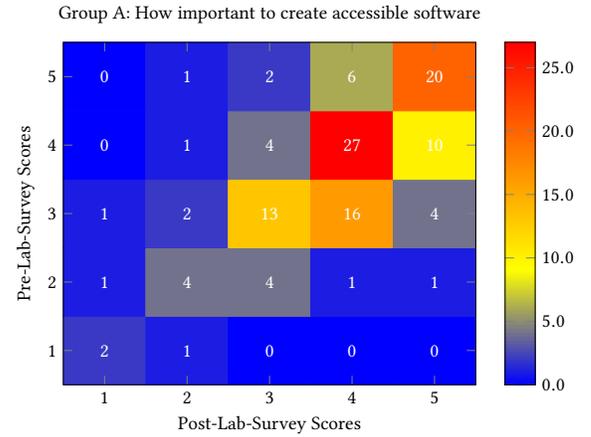

The t-test demonstrates the general impact of each evaluation group on each student group. However, some negative effects can be neutralized by the averaging operation. To better understand the effectiveness of the labs, as shown in Figure~\ref{fig:HM-AJ}, we computed the score transition matrices, $T$, where $T_{ij}$ is the population of students transit from pre-lab-survey score i to post-lab-survey score j. The positive effect of the activity can be read from the lower triangle of the matrix where all entities represent students who used to have low pre-lab-survey scores, but result with higher post-lab-survey score. The diagonal entities represent the students who are not affected by the provided instructional material, and the upper-triangle of the matrix represents the population that are negatively affected by their experiences with the activity.


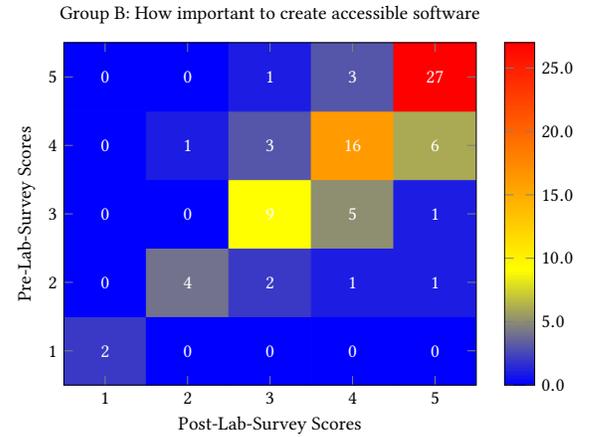
\begin{figure}[h]
 \begin{center}
  \begin{tikzpicture}[scale=.8]
        \begin{axis}[
            view={0}{90},   
            xlabel=Post-Lab-Survey Scores,
            ylabel=Pre-Lab-Survey Scores,
            colorbar,
            colorbar style={
                title=,
                yticklabel style={
                    /pgf/number format/.cd,
                    fixed,
                    precision=1,
                    fixed zerofill,
                },
            },
            title=Group B: How important to create accessible software,
            %
            enlargelimits=false,
            axis on top,
            point meta min=0,
            point meta max=27,
            %
        ]

            \addplot [matrix plot*,
            point meta=explicit,
            nodes near coords,
            every node near coord/.style={white, yshift=-6pt}] file [meta=index 2] {HM-BJ.dat};
        \end{axis}
    \end{tikzpicture}
    	\caption{Score transition heatmap of Group B. The number
in i-th row and j-th column indicates how many students
change their survey score from i to j in Group B (our material).}
  	\label{fig:HM-BJ}
\end{center}
\end{figure}



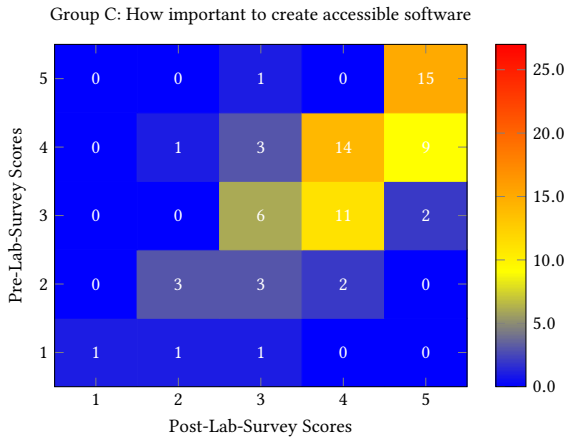
\begin{figure}[h]
 \begin{center}
  \begin{tikzpicture}[scale=.8]
        \begin{axis}[
            view={0}{90},   
            xlabel=Post-Lab-Survey Scores,
            ylabel=Pre-Lab-Survey Scores,
            colorbar,
            colorbar style={
                title=,
                yticklabel style={
                    /pgf/number format/.cd,
                    fixed,
                    precision=1,
                    fixed zerofill,
                },
            },
            title=Group C: How important to create accessible software,
            %
            enlargelimits=false,
            axis on top,
            point meta min=0,
            point meta max=27,
            %
        ]

            \addplot [matrix plot*,
            point meta=explicit,
            nodes near coords,
            every node near coord/.style={white, yshift=-6pt}] file [meta=index 2] {HM-CJ.dat};
        \end{axis}
    \end{tikzpicture}
    	\caption{Score transition heatmap of Group C.The number
in i-th row and j-th column indicates how many students
change their survey score from i to j in Group C (our material with empathy-creating content).}
  	\label{fig:HM-CJ}
\end{center}
\end{figure}


We want to emphasize three observations from the transition matrices. First, Figure~\ref{fig:HM-AJ} shows that although Group A had an overall positive effect on students, its negative effect on students with high pre-lab-survey scores is also significant. Specifically, some students changed their score from 5 (high motivation) in the pre-lab-survey to lower scores in the post-lab-survey. Second, Figure ~\ref{fig:HM-BJ} shows that Group B has most of the population located close to the diagonal, indicating it is truly not impacting students at all score levels (vs positively impacting students at one score level and negatively impacting students at the other score levels. Third, Figure~\ref{fig:HM-CJ} shows that the material included for Group C exhibits a unique educational benefit by positively impacting students who think the topic of accessibility is of very low importance in the pre-lab-survey (score 1 and 2).

To summarize, the primary findings of this research question include: 
\begin{itemize}[noitemsep]
  \item We have a greater than 90\% confidence that both Group A and Group C have an overall positive impact on motivating the students. Group B does not exhibit a significant impact on motivating the students.

  \item The expected impact of Group C is 117\%($\frac{0.37-0.17}{0.17}$) higher than that of Group A.
  \item Group C shows unique power/potential of improving the motivations of students with low starting scores/status. While Group A shows the risk of reducing the motivations of students with high starting scores/status.
  \item The impactfulness of Group C is consistent and robust. The impactfulness of Group A is noisy and unstable.
\end{itemize}

\hangindent=3em \hangafter=1
\textbf{RQ2.} \emph{\RQB} 

To answer this research question, we again compared Group A (existing material) against Group B \& Group C (our created material). We then evaluated the post activity quiz scores for each of these groups to better understand the impact that each set of material had on informing students about the addressed accessibility topic. 



\pgfplotsset{
    boxplot/lower notch/.initial=\pgfutil@empty,
    boxplot/upper notch/.initial=\pgfutil@empty,
    boxplot/notch width/.initial=0.9,
    boxplot/draw/box/.code={%
        \draw[/pgfplots/boxplot/every box/.try]
            (boxplot box cs:\pgfplotsboxplotvalue{lower quartile},0)
            -- (boxplot box cs:\pgfplotsboxplotvalue{lower notch},0)
            -- (boxplot box cs:\pgfplotsboxplotvalue{median},0.5-\pgfplotsboxplotvalue{notch width}/2)
            -- (boxplot box cs:\pgfplotsboxplotvalue{upper notch},0)
            -- (boxplot box cs:\pgfplotsboxplotvalue{upper quartile},0)
            -- (boxplot box cs:\pgfplotsboxplotvalue{upper quartile},1)
            -- (boxplot box cs:\pgfplotsboxplotvalue{upper notch},1)
            -- (boxplot box cs:\pgfplotsboxplotvalue{median},0.5+\pgfplotsboxplotvalue{notch width}/2)
            -- (boxplot box cs:\pgfplotsboxplotvalue{lower notch},1)
            -- (boxplot box cs:\pgfplotsboxplotvalue{lower quartile},1)
            -- cycle
        ;
    },%
    boxplot/draw/median/.code={%
        \draw[/pgfplots/boxplot/every median/.try]
            (boxplot box cs:\pgfplotsboxplotvalue{median},0.5-\pgfplotsboxplotvalue{notch width}/2)
            --
            (boxplot box cs:\pgfplotsboxplotvalue{median},0.5+\pgfplotsboxplotvalue{notch width}/2)
        ;
    },%
        boxplot prepared from table/.code={
        \def\tikz@plot@handler{\pgfplotsplothandlerboxplotprepared}%
        \pgfplotsset{
            /pgfplots/boxplot prepared from table/.cd,
            #1,
        }
    },
    /pgfplots/boxplot prepared from table/.cd,
        table/.code={\pgfplotstablecopy{#1}\to\boxplot@datatable},
        row/.initial=0,
        make style readable from table/.style={
            #1/.code={
                \pgfplotstablegetelem{\pgfkeysvalueof{/pgfplots/boxplot prepared from table/row}}{##1}\of\boxplot@datatable
                \pgfplotsset{boxplot/#1/.expand once={\pgfplotsretval}}
            }
        },
        make style readable from table=lower whisker,
        make style readable from table=upper whisker,
        make style readable from table=lower quartile,
        make style readable from table=upper quartile,
        make style readable from table=median,
        make style readable from table=lower notch,
        make style readable from table=upper notch
}

\pgfplotstableread{
    lw lq med  uq uw  ln  un
    41  58 75 83 100 71 78
    33  66 75 91 100 71 79 
    33  66 83 91 100 79 87
    

    
    
}\datatable

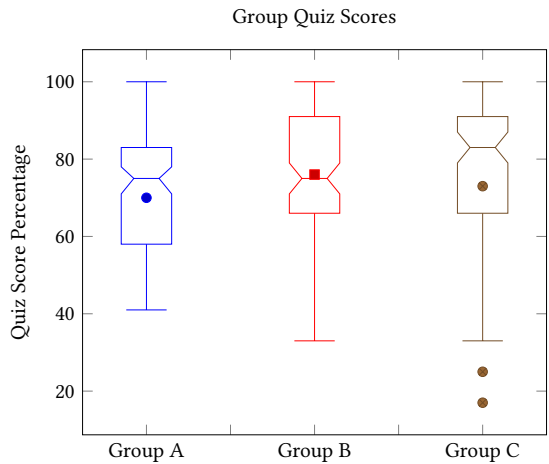
\begin{figure}[h]
 \begin{center}
  \begin{tikzpicture}[scale=.9]
\begin{axis}[
    boxplot/draw direction=y, 
    boxplot/box extend=0.3,
    boxplot/notch width=0.5,
    ylabel=Quiz Score Percentage,
    title=Group Quiz Scores,
    xticklabels={,, Group A, , Group B, ~, Group C} 
]

  \addplot +[
        boxplot prepared from table={
        table=\datatable,
        row=0,
        lower whisker=lw,
        upper whisker=uw,
        lower quartile=lq,
        upper quartile=uq,
        lower notch=ln,
        upper notch=un,
        median=med
    }, boxplot prepared
] coordinates {(2,70)}; 

  \addplot +[
        boxplot prepared from table={
        table=\datatable,
        row=1,
        lower whisker=lw,
        upper whisker=uw,
        lower quartile=lq,
        upper quartile=uq,
        lower notch=ln,
        upper notch=un,
        median=med
    }, boxplot prepared
] coordinates {(4,76)}; 

   \addplot +[
        boxplot prepared from table={
        table=\datatable,
        row=2,
        lower whisker=lw,
        upper whisker=uw,
        lower quartile=lq,
        upper quartile=uq,
        lower notch=ln,
        upper notch=un,
        median=med
    }, boxplot prepared
] coordinates {(6,73)(6,25)(6,17)}; 

\end{axis}
\end{tikzpicture}
    \caption{Student quiz scores for the three evaluation groups.}
  	\label{fig:BP-QuizScores}
\end{center}
\end{figure}


Each student's quiz score is represented by the percentage of their correctly answered questions where each quiz question is equally weighted. Boxplots in Figure~\ref{fig:BP-QuizScores} characterize the score samples of different groups. The boxplot adopts the Tukey style ~\cite{tukey1977exploratory} (\eg the reach of the Whiskers indicates the upper and lower boundaries for outliers). 
Samples outside of the whiskers range are considered as outliers. The notch mark is also applied to boxplotting, indicating the 95\% confidence interval for the median. As demonstrated by student pre-lab-survey data, we can assume that the majority of students have no prior knowledge that could bias them from randomly choosing the correct quiz answers. Then according to the central limit theorem, we can give the following proposition to provide formal definition of the baseline to this research question. The outliers with low quiz scores in Group C likely indicate that a few students merely did not put in reasonable efforts for either understanding the material or for properly completing the quiz.



\begin{prop}
The quiz scores of Students who are given non-effective material follow the normal (Gaussian) distribution with the mean of 0.5.
\end{prop}

We can observe from the Figure~\ref{fig:BP-QuizScores} boxplot that the quiz scores of Group A meets the non-effective material proposition. To further confirm this, we performed a combination of skew test and kurtosis test~\cite{d1971omnibus,royston1992tests} to justify the normality of Group A. The resulting p-value is 0.017, which indicates that Group A is the least affected by the material. Group B and Group C perform equally better than Group A in terms of IQR. However, the median of Group C is significantly better than that of Group B as notches are not overlapping~\cite{krzywinski2014visualizing}. The overlap of the notches does not necessarily rule out a significant difference between two groups, so therefore we can still claim that both Group B and Group C are superior to Group A. 

To summarize, the primary findings of this research question include:

\begin{itemize}[noitemsep]
  \item Both Group B and C exhibit a positive effect on informing students about foundational accessibility principles. 
  \item Group C has a more universal positive effect on students compared with Group B.
  \item Group A (existing material) does not show significant effect on informing students about foundational accessibility principles.  
\end{itemize}

\vspace{2mm}
\hangindent=3em \hangafter=1
\textbf{RQ3.} \emph{\RQC} 

To answer this research question, we performed a comparison of Group B (our material \emph{with no} empathy-creating components) against Group C (our material \emph{with} empathy-creating components). We again used the question `How important is it for you to create accessible software?' from the pre and post-lab-surveys. We began by conducting a t-test, as described in RQ1. We then calculated the t-statistics of the pre-post difference of two groups. We found that the students from Group B and Group C perform differently with respect to this survey question. Group C has a higher post/pre-lab-survey difference than Group B with the p-value of 0.04. This, in correlation to the results in RQ1, indicates that additional empathy-creating material increases student feelings that creating accessible software is important. 

To summarize, the primary findings of this research question include:

\begin{itemize}[noitemsep]

    \item We found that empathy-creating material increases student awareness of the importance of creating accessible software.

    \item The findings of RQ3 further support the observations of RQ1 and RQ2 for the benefits of empathy-creating material in accessibility education.

\end{itemize}


\section{Discussion} 
\label{sec: Discussion}




\subsection{Primary Implications/Findings} 
\label{sec: PrimaryImplications}


Our findings demonstrate that Group C is more consistent at improving student motivation while being less likely to reduce it when compared to Group A. The impact of Group C on student motivation is also more consistent than Group A. This indicates that our material is more effective in not dissuading student motivation on the topic of computing accessibility. This characteristic of not discouraging student motivation regarding the topic of accessibility is crucial. Material should motivate initially uninterested students regarding creating accessible software, but equally as important, not decrease an already interested student's motivation in this topic. 


We found that Group C's empathy-creating material can have an overall positive impact on student motivation. This not only demonstrates the importance and benefits of empathy-creating material in computing accessibility education, but its likely benefits in accessibility education in general.




Group B and C both show a positive impact on informing students about accessibility, while Group C has a more universally positive impact compared to Group B. While we are unable to definitively know the reason for this, we can surmise that since students in Group C were more motivated on the topic of accessibility (we know that Group C increased student motivation), that they paid more attention to the material provided in the activity; thus increasing their comprehension of the topic. This correlates with research that indicates that students learn better and are more engaged when motivated/interested about a topic~\cite{saeed2012motivation, Zyngier_2011, ZYNGIER20081765, jang2008supporting}.




\subsection{Benefit to Adopting Institution}
\label{sec: BenefitInstitution}

The labs offer several benefits to adopting institutions. Adopting instructors will no longer be required to create `one-off' activities for the inclusion of foundational accessibility material in their curriculum. Due to the self-contained nature of the labs, institutions who do not have accessibility experts will not be prohibited from including accessibility activities in their curriculum. This self-contained nature will also limit the amount of preparation time needed by instructors. Additionally, since the labs do not require any special hardware configurations, institutions with limited technical resources will still be able to easily adopt the labs. The demonstrated educational effectiveness of the created labs can also assure that the adopting institution is providing students with robust, educationally effective material. 









%

\subsection{Benefit to Students}
\label{sec: BenefitParticipants}



%


Our labs will enable students to learn about foundational accessibility concepts, both inside and outside of the classroom. Due to their encapsulated nature, students who wish to use our material will not be limited to using them only in the classroom, but may use the material as individual learners as well. Students who use our material will gain foundational knowledge about creating accessible software, with one of the benefits being that they will be more marketable with this knowledge to employers. Additionally, those who may benefit from our material are not only limited to students, but to professional software developers as well. 


\subsection{Limitations and Future Work}
\label{sec: Lessonslearnedandlimitations}

%



Our preliminary evaluations demonstrate the effectiveness of our material in both educating and motivating students about foundational accessibility topics in software development. Future work will include the development and evaluation of additional labs to address other accessibility topics. We will also implement the labs into additional computing and non-computing courses at not only our institution, but at other institutions as well. 

For evaluation consistency, we compared our lab against one set of existing material~\cite{MozillaAccessability_url}. Further work should be done to corroborate our results against other forms of existing accessibility educational material. As we develop more labs, we will test them against additional existing accessibility education resources, providing further confidence in the abilities of our labs and experiential learning structure. Some of the students (33\%) reported having previous experience in computing accessibility. This prior experience could have impacted their feedback/experiences with our labs.

Despite our promising results, there is future work to be conducted and threats to the validity of our study. The are nearly an infinite amount of accessibility topics, and it is impossible to cover all of them in any number of labs. Therefore our labs can only cover a very small portion of possible accessibility issues. We conducted our analysis using primarily first year students. Work should be done to determine if our findings also apply to students of different experience levels. This should be conducted at both the K-12, undergraduate and graduate levels. It would also be interesting to see if our results remain consistent for students with a substantial amount of experience in accessibility.



%


Our labs have an accessibility emulation feature that enables users to experience the software similarly to what a person with the disability would experience. However, despite our best efforts it is unreasonable to expect any software to completely emulate the experiences of someone with the actual disability. 
To provide reasonable confidence that each lab emulated the accessibility experience as closely as possible, we worked with a person who had the specific disability addressed in the lab to determine whether our emulation feature functioned as accurately as possible. For example, we worked with a person who was colorblind for our colorblindness lab and a Deaf/Hard of Hearing student for our hearing lab. 

In evaluating RQ1 (How effective are the labs in motivating students about the importance of accessibility?) We found that Group A \& Group C both have an overall positive impact on motivating students. However, Group B does not exhibit a statistically significant impact on student motivation. Future work should be done to understand why Group B does not have a statistically significant positive impact on student motivation. In answering RQ3, we found that adding empathy material helps students value creating accessible software. The improvement though, is not significant when compared with Group A (existing material). Future work should be done to understand why this is the case.

\section{Related Work} 
\label{sec: RelatedWork}

Existing projects have created accessibility related educational activities and focused on different methods of accessibility education~\cite{lazar2002integrating, bohman2012teaching, burgstahler2015universal, Poor:2012:NUL:2160547.2160548, MozillaAccessability_url}. However, to our knowledge none: (I) Have been thoroughly evaluated to determine their educational effectiveness (II) Offer a complete experiential learning experience, proving all instructional material (III) Are hosted and do not require the installation of any software (IV) Contain empathy building material. Teaching accessibility in computing courses has been a significant challenge in higher education \cite{Kawas:2019:TAD:3287324.3287399, bohman2012teaching}. While some institutions have developed entire courses or degrees devoted towards the topic of accessibility, our work focuses on creating easily adoptable material that can readily integrate into existing curriculum. 




Kane \etal\cite{kane2007engaging} described an initial study where a web programming course used pedagogical techniques drawn from architecture and industrial design support students in empathizing with users with disabilities. Initial observations indicate that this approach is effective in encouraging accessible design practices. This work differs from ours in that it focused on encouraging students from a design perspective, and did not provide a complete set of hands-on activities such as those in our work. Additionally, the evaluations were conducted on a much smaller scale (17 vs. \TotalStudentsEvaluation students).

There are also accessibility teaching materials available online. For example, the `Teach Access Tutorial' provides developers and designers with a set of lessons and exercises that teach basic accessible web development practices \cite{teach_access_tutorial_url}. Additional teaching resources are compiled by \emph{AccessComputing}\cite{access_computing_url}, which is an alliance that supports students with disabilities learn computing. AccessComputing focuses on making computing courses accessible to students with disabilities, and also on supporting instructors teaching about accessibility. For example, AccessComputing shares curriculum resources \eg educational components that teach students and developers how to create accessible mobile applications \cite{El-Glaly:2018:AEM:3211407.3182184}. To our knowledge, no existing material provides a complete educational experience (experiential activity, lecture slides, \etc) that have been evaluated to demonstrate their educational effectiveness as we have done with our Accessibility Learning Labs.

Lewthwaite \etal\cite{lewthwaite2016exploring} identified several of the challenges for teaching and learning accessibility in computing education. This work contends that accessibility education in computing presents a set of unique and challenging characteristics. Our work differs in that we do not focus on identifying specific challenges in computing accessibility education, but focus in presenting and evaluating a set of unique experiential educational materials.


Educators have integrated accessibility into existing courses such as web design \cite{Rosmaita:2006:AFN:1121341.1121426, Wang:2012:HPA:2380552.2380568}, HCI \cite{petrie2006inclusive, Poor:2012:NUL:2160547.2160548}, and software engineering courses \cite{ludi2007introducing} using various pedagogical methods such as lectures \cite{Wang:2012:HPA:2380552.2380568}, programming activities \cite{El-Glaly:2018:AEM:3211407.3182184}, and projects \cite{Poor:2012:NUL:2160547.2160548, ludi2007introducing, lazar2011using}. Educators found that when students interact with individuals with disabilities, \eg project stakeholders, they better understand and apply accessibility principles in their work \cite{ludi2007introducing, lazar2011using}. Similarly, students who watched videos for individuals with disabilities \cite{putnam2016best} and older adults \cite{carmichael2007efficacy}, or were required to use assistive technology \eg screen readers \cite{Harrison:2005:OES:1047344.1047368} were found to be more aware of the needs of the diverse base of users \cite{Putnam:2015:TAL:2700648.2811365}.







Industry has partnered with academia and advocates for people with disabilities in an initiative known as \emph{Teach Access}~\cite{TeachAccess_url}. A goal of Teach Access is to improve accessibility education in higher education \cite{Lawrence:2017:TAP:3017680.3022392}. Despite these efforts and previously published work, including accessibility in computing courses is still an individual effort that is driven by faculty who have experience in accessibility or a related field \eg HCI \cite{Putnam:2015:TAL:2700648.2811365}, constituting only approximately 2.5\% of instructors \cite{Shinohara:2018:TAS:3159450.3159484}. 
Recent interviews and surveys indicate that computing instructors have the desire to integrate accessibility-related topics in their courses, however they frequently lack access to teaching materials to use in their courses \cite{Shinohara:2018:TAS:3159450.3159484, Kawas:2019:TAD:3287324.3287399}. We address this problem by creating instructional resources that are easy to integrate into existing courses, with defined learning objectives. 

Our labs adhere to experiential learning principles, which have been shown to be beneficial to computing education~\cite{botelho2016kolb, kolb2005learning, kiili2005digital}. Experiential learning provides a complete learning experience for the student, one where they both understand the concept behind an idea and interactively learn about it
~\cite{boud2013reflection}. Within the context of experiential learning, different activities have been employed by instructors such as exercises
~\cite{hamer2000additive}, projects~\cite{carter1986dear}, simulations~\cite{silberman2007handbook}, and role-playing~\cite{nestel2007role}. Experiential learning, compared to traditional teaching approaches such as lectures, has been demonstrated to be more engaging for students~\cite{Laird:2016:ESE:2889160.2889205}, and supports student retention of information~\cite{specht1991differential, hawtrey2007using}. Examples of experiential learning in computing education include teaching software engineering using interactive tutorials~\cite{Krusche:2017:ILI:3013499.3013513} and software estimation using LEGOs~\cite{Laird:2016:ESE:2889160.2889205}.

There have been a large number of previous works that have examined best practices for motivating students in computing education. These focus on a wide-range of topics such as general computing and cybersecurity to how to best motivate students in an online instructional format~\cite{keller1995motivation, guzdial2006imagineering, lee2000student, mahle2011effects, robey2006student}. Our work differs in that we specifically focus on computing accessibility education, while additionally seeking to determine the specific impact of empathy-creating material in computing accessibility education for both motivating and informing students.

\section{Conclusion}
\label{sec: Conclusion}
This work demonstrates the positive impact of experiential learning in computing accessibility education, specially through the use of our publicly available \emph{Accessibility Learning Labs} (ALL). Our primary findings demonstrate: (I) The potential of the proposed experiential learning format and that the labs are effective in motivating students about the importance of accessibility (II) The proposed material is effective in informing students about foundational accessibility topics (III) Empathy-creating material is demonstrated to be a beneficial component in accessibility education, supporting students in placing a higher value on the importance of creating accessible software. Created labs and project materials are publicly available on the project website: \ALLURL





\section*{Acknowledgements}

This material is based upon work supported by the United States National Science Foundation under grant \#1825023.


\clearpage
\balance
\bibliographystyle{ACM-Reference-Format} 
\bibliography{refs}

\end{document}